# Estimation of surface tension in nuclei

N.G. Goncharova
Moscow State University, Department of Physics, 119991 Moscow, Russia.
n.g.goncharova@gmail.com; n.g.goncharova@physics.msu.ru

**Abstract:** Coefficients of surface tension for even-even nuclei were estimated using their dependence of nuclear rigidity. The values of nuclear rigidity were obtained owing to its connection to the mean squared deformations. The correlation of estimated surface tension coefficients with the data on nuclear shapes was revealed.

## 1. Introduction

Simultaneously with the creation of the liquid drop model [1] the concept of nuclear surface tension was adopted. The model explained the experimental data on the dependence of the nuclear binding energy on the numbers of nucleons and protons.

The value of nuclear surface tension determines the ratio of the second (surface) term to the size of nuclear surface. The connection between the nuclear surface tension $\sigma$ and the rigidity C of the nucleus was already shown at the dawn of nuclear physics [2,3]:

$$C_\lambda = (\lambda-1)(\lambda+2)R^2\sigma - \frac{3(\lambda-1)Z^2 e^2}{2\pi(2\lambda+1)R}. \qquad (1.1)$$

In (1) $\lambda$ denotes the multipolarity of the collective surface vibrations. The rigidity determines the contribution of potential energy into Hamiltonian of nuclear collective vibrations:

$$\hat{H}_{coll.vib.} = \frac{1}{2D}\sum_{\lambda,m}|\hat{b}_{\lambda,m}|^2 + \frac{C}{2}\sum_{\lambda,m}|\hat{a}_{\lambda,m}|^2. \qquad (1.2)$$

In (2) $\hat{a}$ denotes the operator of collective vibrations' coordinate with the multipolarity $\lambda$ which determines the dynamics of the nuclear surface:

$$R(r,\theta,\varphi) = R_0[1+\sum_{\lambda,m}a^*_{\lambda m}Y_{\lambda m}(\theta,\varphi)] \qquad (1.3)$$

The study of nuclear spectra and properties has shown the dominant role of quadrupole vibrations in the Hamiltonian (2) for the majority of nuclei.

The value of mean squared coordinate of quadrupole vibrations (mean squared deformation) depends on the nuclear rigidity C and the number of phonons in the nuclear state (see e.g. [4]):

$$\beta_N^2 = \left\langle N, J, M \left| \sum_\mu |\hat{a}_\mu|^2 \right| N, J, M \right\rangle = \frac{\hbar\omega}{2C}(2N + 5). \quad (1.4)$$

In (4) $\hbar\omega$ denotes the energy of the quadrupole phonon. It approximately corresponds to the energy of the lowest $2^+$ state of even-even nucleus.

## 2. Rigidities and coefficients of surface tension

For the estimation of nuclear rigidities in ground states with N=0 phonons we can use the relation

$$C \cong \frac{5E(2^+)}{2\beta_{N=0}^2}. \quad (2.1)$$

The experimental investigation of $2^+ \to 0$ transitions probabilities in the even-even nuclei (reviewed in [5]) permits us to perform the estimation of the mean squared deformations $\beta$ for the majority of even-even atomic nuclei. Data [5] on the mean squared deformations were used for calculation of nuclear rigidities in [6, 7].

The comparison of the obtained data with the nuclear shell structure has demonstrated the impact of the shells' filling on nuclear rigidities and nuclear charge densities as well. For instance, for the even-even calcium isotopes the maximal values of rigidities C were obtained for $^{48}$Ca and $^{40}$Ca, with $C(^{48}\text{Ca}) > C(^{40}\text{Ca})$. For the same nuclei, the values of coefficients $r_0$ (in the formula for nuclear radii $R = r_0 A^{1/3}$) reach their minima. (It should be mentioned that the estimations of $r_0$ were based on the charge nuclear radii data [8] ).

The correlations between the nuclear rigidities' maxima/ minima and $r_0$ are the results of the surface tension's impact on the properties of a nucleus. High rigidity corresponds to high values of surface tension σ which caused high pressure on the nuclear sphere, according to the "classic" Laplace's formula

$$p = \frac{2\sigma}{R}. \quad (2.2)$$

The connection of nuclear surface tension and nuclear rigidity (1.1) allows us to estimate the surface tension coefficients σ for even-even nuclei. The calculations were performed for all stable (or near-stability) nuclei from A=12 up to A=210.

In the Fig.1 the distribution of the surface tension coefficients in even calcium isotopes is represented together with the corresponding parameters $r_0$. The influence of filling of the neutron subshell $(1f_{7/2})_n^8$ on the nuclear charge density is revealed.

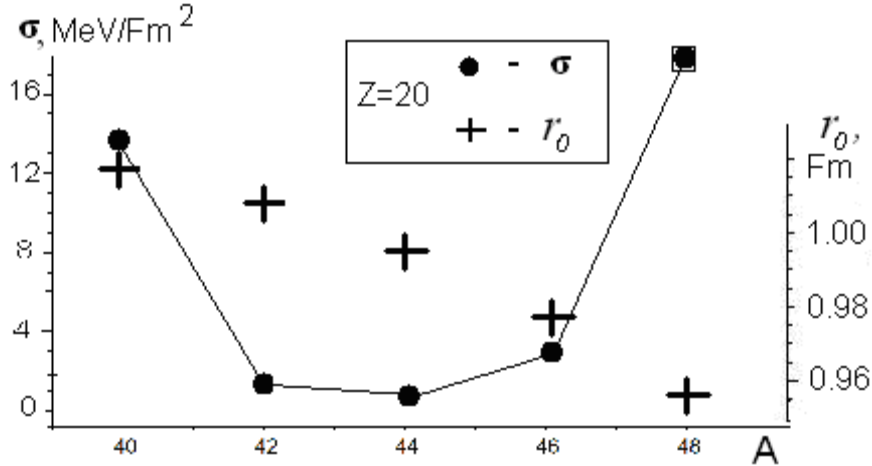

*Fig.1. The surface tension coefficients and parameters $r_0$ in even calcium isotopes.*

In the paper [9] the dependence of nuclear charge radii on the shell structure was demonstrated on the results of experimental definition of $R_{ch}$ for Ar, Ca, K, Ti and Cr isotopes. For all these nuclei the minima of the mean squared radii were localized at the neutron number N=28. It should be mentioned that the parameter $r_0 = R_{ch} \cdot A^{-1/3}$ goes through the minimum at N=28 for all the investigated *sd*-shell nuclei.

In comparison with the neutron subshells, the filling of the proton subshells is less likely to lead to the high surface tension in light nuclei. In the Fig.2 the dependence of the surface tension coefficients on the proton numbers Z is shown for the even nuclei with the neutron number N=28.

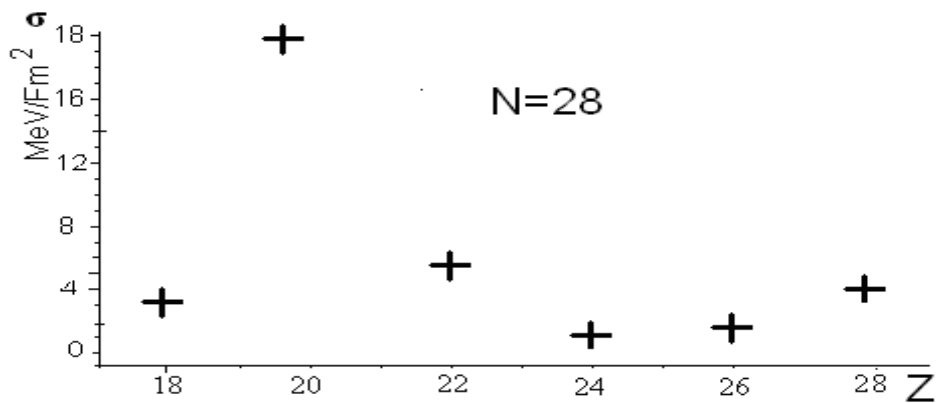

*Fig.2. Surface tension in nuclei with neutron number N=28 and even proton numbers Z.*

The most striking result of the performed calculations for the surface tension coefficients is a very high variance of their values. The Fig.3 shows the plot of the surface tension coefficients for the nuclei with nucleons numbers 10<A<100. The results of estimation of rigidities and the surface tension are represented in the Table 1. The numbers of the nucleons and the corresponding values of rigidity C (MeV) and surface tension coefficients (MeV/Fm$^2$) are listed.

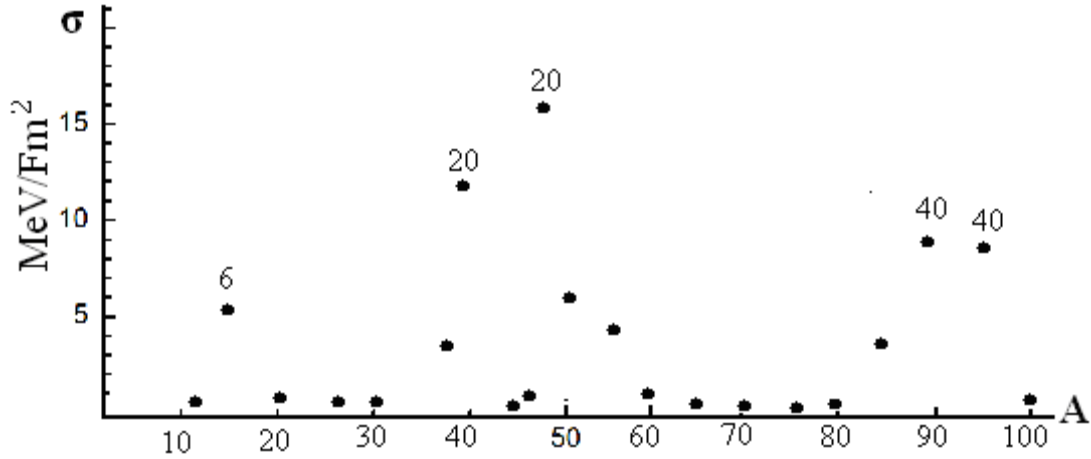

Fig.3. Distribution of σ in 10<A<100 nuclei. Numbers on the plot – the values of Z.

Table 1

Numbers of nucleons (A), protons (Z) and neutrons (N) and corresponding C (MeV) and σ(MeV/Fm$^2$). (12≤A≤66)

| A | 12 | 14 | 16 | 18 | 20 | 22 | 24 | 26 | 28 | 30 | 32 | 34 | 36 | 38 |
|---|----|----|----|----|----|----|----|----|----|----|----|----|----|----|
| Z | 6 | 6 | 8 | 8 | 10 | 10 | 12 | 12 | 14 | 14 | 16 | 16 | 16 | 18 |
| N | 6 | 8 | 8 | 10 | 10 | 12 | 12 | 14 | 14 | 16 | 16 | 18 | 20 | 20 |
| C | 33 | 135 | 30.5 | 39.3 | 7.7 | 10.1 | 9.4 | 19.5 | 27 | 56 | 57 | 84 | 291 | 204 |
| σ | 1.42 | 5.48 | 4.57 | 1.38 | 0.34 | 0.42 | 0.4 | 0.71 | 0.91 | 1.65 | 1.6 | 2.19 | 6.94 | 4.68 |

| A | 40 | 42 | 44 | 46 | 48 | 50 | 52 | 54 | 56 | 58 | 60 | 62 | 64 | 66 |
|---|----|----|----|----|----|----|----|----|----|----|----|----|----|----|
| Z | 20 | 20 | 20 | 20 | 20 | 22 | 24 | 26 | 28 | 28 | 28 | 28 | 30 | 30 |
| N | 20 | 22 | 24 | 26 | 28 | 28 | 28 | 28 | 28 | 30 | 32 | 34 | 34 | 36 |
| C | 645 | 62 | 45 | 144 | 853 | 140 | 72 | 93 | 226 | 108 | 78 | 75 | 42 | 55 |
| σ | 13.7 | 1.6 | 1.2 | 3.26 | 17.96 | 5.47 | 1.67 | 2.03 | 4.34 | 2.40 | 1.83 | 1.75 | 1.19 | 1.38 |

The high values of σ for 40<A<50 nuclei correspond to calcium isotopes. For the nuclei with 48<A<90 the maximal values of σ were obtained for Kr-84 (C~100 MeV, σ~ 4 MeV/Fm$^2$) and Sr-88 (C~330MeV, σ~5.3 MeV/Fm$^2$). In all remaining nuclei in this area of A the values of σ oscillate within the limits (0.8÷2) MeV/Fm$^2$.

The results of the estimation of rigidities and surface tension for nuclei with A>66 are represented in the table 2 (68≤A≤150).

Table 2

Numbers of nucleons (A), protons (Z), neutrons (N), rigidities C(MeV) and σ(MeV/Fm$^2$ ).
(68≤A≤150)

| A | 68 | 70 | 72 | 74 | 76 | 78 | 80 | 82 | 84 | 86 | 88 | 90 | 92 | 94 | 96 |
|---|---|---|---|---|---|---|---|---|---|---|---|---|---|---|---|
| Z | 30 | 32 | 32 | 32 | 34 | 34 | 34 | 36 | 36 | 36 | 38 | 40 | 40 | 40 | 40 |
| N | 38 | 38 | 40 | 42 | 42 | 44 | 46 | 46 | 48 | 50 | 50 | 50 | 52 | 54 | 56 |
| C | 64 | 52 | 36 | 19 | 15 | 21 | 31 | 48 | 99 | 137 | 334 | 685 | 222 | 282 | 683 |
| σ | 1.5 | 1.3 | 0.8 | 0.8 | 0.8 | 0.9 | 1.0 | 1.3 | 4.0 | 2.6 | 5.3 | 10.0 | 3.7 | 4.4 | 9.65 |

| A | 96 | 98 | 100 | 102 | 104 | 106 | 108 | 110 | 112 | 114 | 116 | 118 | 120 | 122 |
|---|---|---|---|---|---|---|---|---|---|---|---|---|---|---|
| Z | 42 | 42 | 42 | 44 | 44 | 46 | 46 | 48 | 48 | 48 | 50 | 50 | 50 | 50 |
| N | 54 | 56 | 58 | 58 | 60 | 60 | 62 | 62 | 64 | 66 | 66 | 68 | 70 | 72 |
| C | 66 | 69 | 25 | 21 | 12 | 24 | 18 | 53 | 45 | 39 | 259 | 252 | 253 | 266 |
| σ | 1.6 | 1.6 | 1.0 | 1.0 | 0.9 | 1.2 | 1.1 | 1.6 | 3.3 | 1.4 | 3.9 | 3.8 | 3.8 | 3.9 |

| A | 124 | 126 | 128 | 130 | 132 | 134 | 136 | 138 | 140 | 142 | 144 | 146 | 148 | 150 |
|---|---|---|---|---|---|---|---|---|---|---|---|---|---|---|
| Z | 50 | 52 | 52 | 52 | 54 | 56 | 56 | 56 | 58 | 58 | 60 | 60 | 62 | 62 |
| N | 74 | 74 | 76 | 78 | 78 | 78 | 80 | 82 | 82 | 84 | 84 | 86 | 86 | 88 |
| C | 312 | 71 | 100 | 150 | 84 | 58 | 127 | 412 | 387 | 98 | 114 | 49 | 68 | 22 |
| σ | 4.4 | 1.7 | 2.0 | 2.5 | 1.8 | 1.6 | 6.9 | 5.35 | 5.1 | 2.0 | 2.2 | 1.5 | 5.96 | 1.24 |

The most striking correlation between the high values of surface tension and the low values of $r_0$ parameters was revealed for zirconium isotopes (Fig.4)

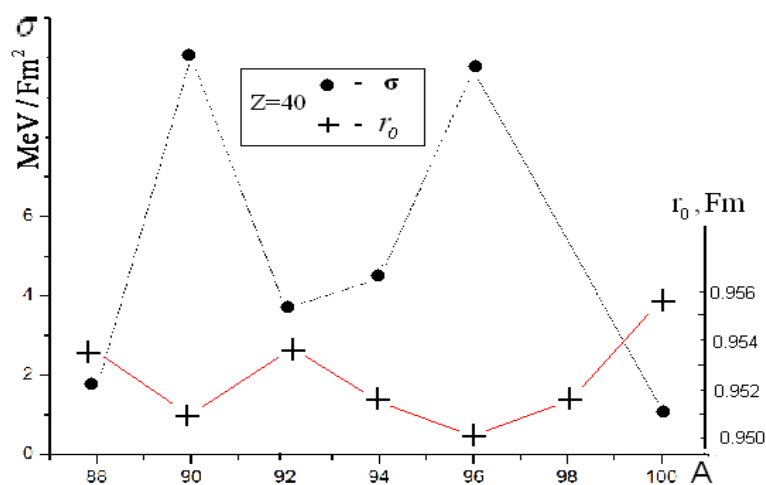

*Fig.4. Surface tension coefficients σ and $r_0$ for even Zirconium isotopes.*

Zr-90 nucleus has 10 neutrons in the last sub-shell, Zr-96 nucleus has 16 neutrons in the surface subshell $(1g_{9/2})_n^{10}(2d_{5/2})_n^6$ . The complementary subshell with 6 neutrons in Zr-96 leads to diminishing of $r_0$ parameter in comparison with Zr-90, and to a higher protons density. The correlations between the high values of σ and low values of $r_0$ demonstrate the impact of the neutron shell structure on nuclear properties.

The distribution of σ in nuclei with 100<A<150 is shown in the Fig.5. The values of surface tensions for nuclei with 100≤ A<110 are low and do not exceed 2 $MeV/Fm^2$. In the area of 112≤A<130 the highest values of σ belong to the even-even isotopes of tin Sn-122 (3.9 $MeV/Fm^2$) and Sn-124 (4.4 $MeV/Fm^2$). At 130≤A<150 the maximal surface tension coefficients correspond to Ba-136 (Z=56) and Sm-148 (Z=62).

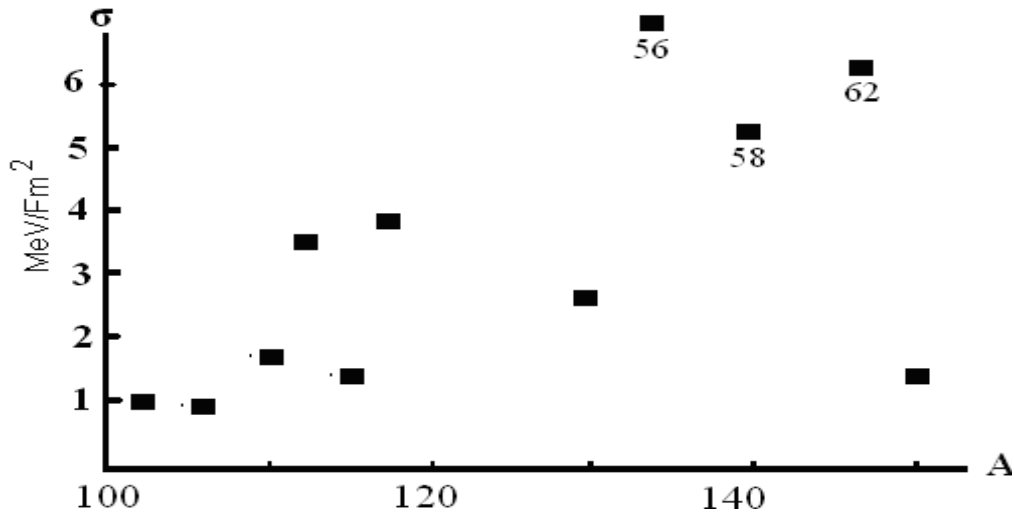

*Fig.5. Distribution of σ in 100<A<150 nuclei. Numbers on the plot are the values of Z.*

The results of the estimations of the nuclear surface tension for heavy nuclei 152≤A≤210 are listed in the Table 3.

For the nuclei with 152≤A<198 the rather low values of surface tension are typical: $1.0\,MeV/Fm^2 \leq σ < 1.8\,MeV/Fm^2$. Low surface tension and high Coulomb's forces in this area lead to the deviation of the nuclear shape from a spheroid.

In the even isotopes of Mercury the surface tension grows with the numbers of neutrons in the last subshell from σ=1.8 $MeV/Fm^2$ (A=196, N=116) up to σ=3.3 $MeV/Fm^2$ (A=204, N=124).

The highest values of the surface tension coefficients have been achieved for the even isotopes of lead (Fig.6). Here the filling of the external proton subshell leads to a sharp rise of surface tension: from σ ≈3.3 $MeV/Fm^2$ for Hg-204 up to the σ ≈7.5 $MeV/Fm^2$ for Pb-206 with the same number of neutrons N=124.

Among all the investigated nuclei the highest value of surface tension belongs to Pb-210 (σ ≈ 33.9 $MeV/Fm^2$).

Table 3

Numbers of nucleons (A), protons (Z), neutrons (N), rigidities C(MeV) and σ(MeV/Fm$^2$ ).
(152≤A≤210)

| A | 152 | 154 | 156 | 158 | 160 | 162 | 164 | 166 | 168 | 170 | 172 | 174 | 176 | 178 | 180 |
|---|---|---|---|---|---|---|---|---|---|---|---|---|---|---|---|
| Z | 62 | 62 | 64 | 64 | 66 | 66 | 66 | 68 | 68 | 68 | 70 | 70 | 72 | 72 | 72 |
| N | 90 | 92 | 92 | 94 | 94 | 96 | 98 | 98 | 100 | 102 | 102 | 102 | 104 | 106 | 108 |
| C | 3 | 1.7 | 2 | 1.6 | 1.9 | 1.7 | 1.5 | 1.7 | 1.7 | 1.7 | 1.8 | 1.8 | 2.5 | 3 | 3 |
| σ | 1.0 | 1.0 | 1.1 | 1.0 | 1.1 | 1.08 | 1.07 | 1.12 | 1.12 | 1.1 | 1.15 | 1.14 | 1.2 | 1.2 | 1.2 |

| A | 182 | 184 | 186 | 188 | 190 | 192 | 194 | 196 | 198 | 200 | 202 | 204 | 204 | 206 | 208 | 210 |
|---|---|---|---|---|---|---|---|---|---|---|---|---|---|---|---|---|
| Z | 74 | 74 | 74 | 76 | 76 | 78 | 78 | 80 | 80 | 80 | 80 | 80 | 82 | 82 | 82 | 82 |
| N | 108 | 110 | 112 | 112 | 114 | 114 | 116 | 116 | 118 | 120 | 122 | 124 | 122 | 124 | 126 | 128 |
| C | 4 | 5 | 6 | 11 | 15 | 19 | 40 | 53 | 91 | 95 | 160 | 232 | 1192 | 1940 | 3310 | 3970 |
| σ | 1.3 | 1.3 | 1.3 | 1.4 | 1.4 | 1.5 | 1.6 | 1.8 | 2.1 | 2.2 | 2.7 | 3.3 | 11.3 | 17.5 | 28.7 | **33.9** |

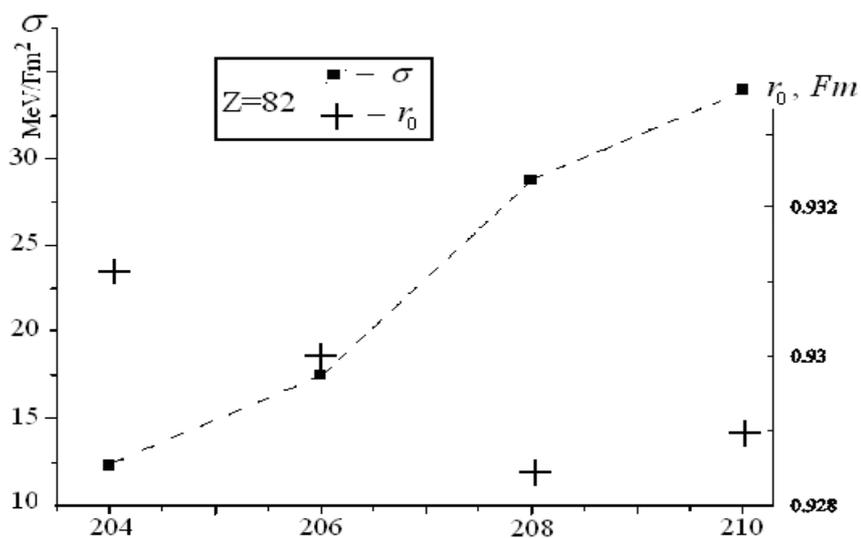

*Fig.6. Coefficients of surface tension **σ** and r$_0$ in the Lead isotopes.*

## 3. Discussion and summary

The represented results of the surface tension calculations should be considered as very rough estimations. The reliability of numerical values of the σ coefficients is influenced by the following factors:

1) The estimations of the surface tension in even-even nuclei were based on the connections between σ coefficients and rigidities of these nuclei. Only quadruple vibrations were taken into account , although the vibrations of other multipolarities also make some contributions into nuclear rigidity;

2) The preciseness of the mean squared deformations $\beta$ [5] does not exceed about 15÷20 %;

3) In the calculations of the surface tension coefficients only the values of charge radii of nuclei were used since for the majority of the nuclei the distributions of the masses are not known yet.

The obtained estimations of the surface tension show great differences in its values for different nuclei. The diversity of the surface tension coefficients is rather high: from about 0.8÷1.2 $\text{MeV}/\text{Fm}^2$ up to $\sigma \approx 34$ $\text{MeV}/\text{Fm}^2$. The surface tension in nuclei is highly influenced by the shell structure, especially of the neutron subshells near the surface. The competition between the surface tension and Coulomb's forces determines the deviations of nuclear shapes from a spheroid.

Author is very grateful to E.Arakelyan for helpful discussion.